\documentclass[prl,twocolumn]{revtex4}

\newcommand{\beq}{\begin{equation}}
\newcommand{\eeq}{\end{equation}}
\newcommand{\ket}[1]{\left| {#1} \right>}
\newcommand{\C}{^C\!}

\usepackage{graphics}
\usepackage{amssymb}

\begin{document}

\title{Fast fault-tolerant filtering of quantum codewords}

\author{Andrew M. Steane}
\affiliation{Centre for Quantum Computation, Department of Atomic
and Laser Physics,\\ Clarendon Laboratory, Parks Road, Oxford, OX1
3PU, England}

\date{\today}

\begin{abstract}
The stabilization of a quantum computer by repeated error correction
can be reduced almost entirely to repeated
preparation of blocks of qubits in quantum codeword states. These
are multi-particle entangled states with a high degree of symmetry.
The required accuracy can be achieved by measuring parity checks, using
imperfect apparatus, and rejecting states which fail them. 
This filtering process is considered for $t$-error-correcting codes with $t>1$.
It is shown how to exploit the structure of the codeword and the check matrix,
so that the filter is reduced to a minimal
form where each parity check need only be measured once, not $> t$ times
by the (noisy) verification apparatus. This both raises the noise threshold
and also reduces the physical size of the computer. A method based
on latin rectangles is proposed, which enables the most parallel version
of a logic gate network to be found, for a class of networks including
those used in verification. These insights allowed the noise threshold 
to be increased by an order of magnitude.
\end{abstract}


\maketitle

The concept of quantum computing has given fundamental insights
into the laws of Nature and promises powerful new computing
capability, beyond the range of any other type of computer, if it
can be realized in practice
\cite{00:Bennett,98:Steane,Bk:Nielsen}. A central consideration in
both these aspects is the intrinsic sensitivity of quantum
processes to the inevitable imperfections in physical devices.
There exist protocols based on quantum error correction (QEC)
\cite{95:Shor,96:SteaneA,96:Calderbank,96:SteaneB}
which allow successful quantum computing in the presence of
low-level noise throughout the computer at all times---this is
called fault-tolerance. A set of ideas which allow QEC to be
fault-tolerant were put forward by Shor \cite{96:Shor,97:Preskill,98:Preskill}. There followed
further insights which generalized or improved the speed and
space-efficiency of the methods
\cite{96:DiVincenzo,97:SteaneA,98:GottesmanA,99:SteaneB,99:GottesmanB}.

The noise level which can be tolerated in QEC has been estimated by 
analysis of the networks of operations
involved \cite{97:Preskill,98:Preskill,97:Zalka,98:Aharonov}.
In \cite{03:Steane} the present author
analysed QEC networks which included several further improvements in efficiency, and found
that the tolerated noise could be considerably higher than was previously possible.
This letter presents the basic insights which generated these efficiency improvements.

This work is significant not only to the practical task of building a quantum computer,
but also to two other areas. First, it contributes to our understanding of the
thermodynamics of controlled entangled systems.
This subject has only been explored a little up till now, but has
revealed some striking behaviour; for example
fault-tolerant (FT) methods lead to a phase transition in a set of
qubits stablized by QEC, in which the order parameter is related
to the size of clusters of qubits whose entanglement is preserved
at finite temperature \cite{00:Aharonov,02:Dennis}. The present work extends the
range of possible transition temperatures. Secondly, basic insights into quantum network
constructions improve our understanding of quantum processing methods in
general, extending their use both in quantum algorithms and in
realizing physical effects which exploit entanglement for other
purposes.

The main result of FT QEC is to achieve a logical
error rate of $ O(\epsilon^{t+1})$ per logical operation followed
by recovery, where $\epsilon$ is the imprecision or noise of the
elementary operations on physical qubits or per time-step for
resting qubits, when a $t$-error correcting quantum code is
employed. In order to do this, a combination of well-chosen
code structure, network construction, and repetition is employed. In
this letter I exploit the first two ingredients so as to avoid
the third. I also show how to arrange the relevant networks using
a method based on latin rectangles so that they
are as parallel as possible and hence require a minimal number of time-steps.

This study assumes the most efficient way 
to achieve FT QEC
known to the author, as follows. Ancillary qubits are
prepared in quantum codeword
states (e.g. the encoded logical zero state $\ket{0}_L$), then they are
coupled them to the `data' qubits which store the (encoded)
logical information, then the ancilla are measured and 
the classical information obtained is decoded by classical processing
in order to deduce the corrective operation to be applied to the data \cite{97:SteaneA}.
This method works for all CSS codes; these are important because they include
good codes and allow relatively simple, and therefore robust, FT gate
constructions \cite{99:SteaneB,96:Shor,98:GottesmanA,99:GottesmanB}.
The method relies on the following concepts. First,
the properties of the encoding lead to the correct movement of error
information from the data qubits to the measurement apparatus
applied to the ancilla, without extracting any of the
quantum information stored in the data. Secondly, the ancilla
needs only to be checked for one type of error (either
$\sigma_x \equiv X$ or $\sigma_z \equiv Z$; I assume $X$ here),
because only one type propagates from ancilla to data when the
two are coupled (when the desired propagation of $Z$
errors from data to ancilla takes place, then $X$ errors propagate
in the other direction, and vice versa). The ancilla errors which are
not detectable by this verification remain in the ancilla and render the deduced
information concerning data errors (i.e. the syndrome) unreliable; the third ingredient
is to use several independently prepared ancillas to extract the
same syndrome, and take a majority vote. These methods are typically
analysed under the assumption that
different gates in the network, and qubits at different
positions or times, fail independently with
probability $\epsilon$. The degree to which this assumption
can be relaxed without significantly affecting the results
is an area of active investigation \cite{98:Aharonov,01:Alicki,03:Steane,01:SteaneA}.

Let the processes which cause imperfection be called
`failures' and the resulting imperfections in the state of the qubits
be called `errors'. Each error $e$ is a tensor product of Pauli operators.
Let $P_{e}(\epsilon)$ be the probability that 
the ancilla's state differs from the desired state by $e$. We define the error
to be {\em uncorrelated} if $P_e$ satisfies
  \[
P_e(\epsilon) = a_e \epsilon^s \; | \;  s \ge w_e \mbox{ for } w_e \le t; \; s >
t \mbox{ for } w_e > t,     
  \]
and we define $e$ to be {\em correlated} otherwise.
Here $w_e$ is the weight of $e$ (the number of qubits it affects).
The coefficients $a_w$ (which depend on the code
and the networks) should not be unreasonably large.

The central resource in the QEC protocol is
an ancilla prepared in an approximation to $\ket{0}_L$ having
uncorrelated $X$ errors. If this is available,
then after coupling it to the data (in order to obtain the
syndrome of the data) the probability of finding
uncorrectable errors in the data scales in the right way, and the
main result of FT QEC theory applies.

Ancilla verification is achieved by measuring
all those observables $M$ in the stabilizer of $\ket{0}_L$ which
anticommute with $X$ errors (and therefore whose measured
eigenvalues reveal the presence of $X$ errors). The problem is
that a failure of $<w$ of these measurements might allow an $X$
error of weight $w$ to go undetected (see figure 1a). This was avoided in Shor's
and subsequent work by repeating the measurements $t+1$ or more
times. This solution is costly in noise tolerance or computer
size or both because for a large computation $t \gg 1$.

I will now show how to avoid the repetition in the case of an
arbitrary CSS code. 

The ancilla is to be placed in the codeword state $\ket{0}_L = \sum_u \ket{u \in
C}$ where $C$ is the classical code on which the CSS code is
based. When 
an approximation to $\ket{0}_L$ is prepared by any means, 
there are typically many locations in the preparation 
where a single failure results in a
high-weight error in the output state; I assume the worst
case that the preparation leaves an error of any non-zero weight with probability
proportional to $\epsilon$. The subsequent measurements of observables $M$
in the $Z$ part of the stabilizer (equivalently, of parity checks
which $\ket{0}_L$ ought to satisfy) must satisfy two conditions:
(1) no correlated $X$
errors are introduced and (2) all correlated $X$ errors are
detected with high enough probability: the probability that an error of
weight $w$ goes undetected must scale as $\epsilon^{s \ge (w-1)}$.
Condition (1) is
guaranteed by the fact that $X$ and $Z$ errors propagate
differently, so that a network of controlled-gates to store $X$
parity information into one verification bit will cause $Z$- but
not $X$-error-propagation around the ancilla bits.
I will prove that a (properly
constructed) noisy network also satisfies (2), where the network
requires only a single measurement of each check, and a single
physical qubit to accumulate each logical check bit, therefore
it is minimal and QEC is rapid.

We require that failures of low weight in the verification network only
cause ancilla errors of low weight to be `missed'. The non-trivial situation
is when at least one failure occurs in the
preparation and at least one in the verification; the case $t=1$
(single-error-correcting code) is trivial since this is
already a second-order process; this is why the question has not
arisen in discussions of the $[[7,1,3]]$ code and its
concatenations.

\begin{figure}[ht]
\centerline{\resizebox{!}{3 cm}{\includegraphics{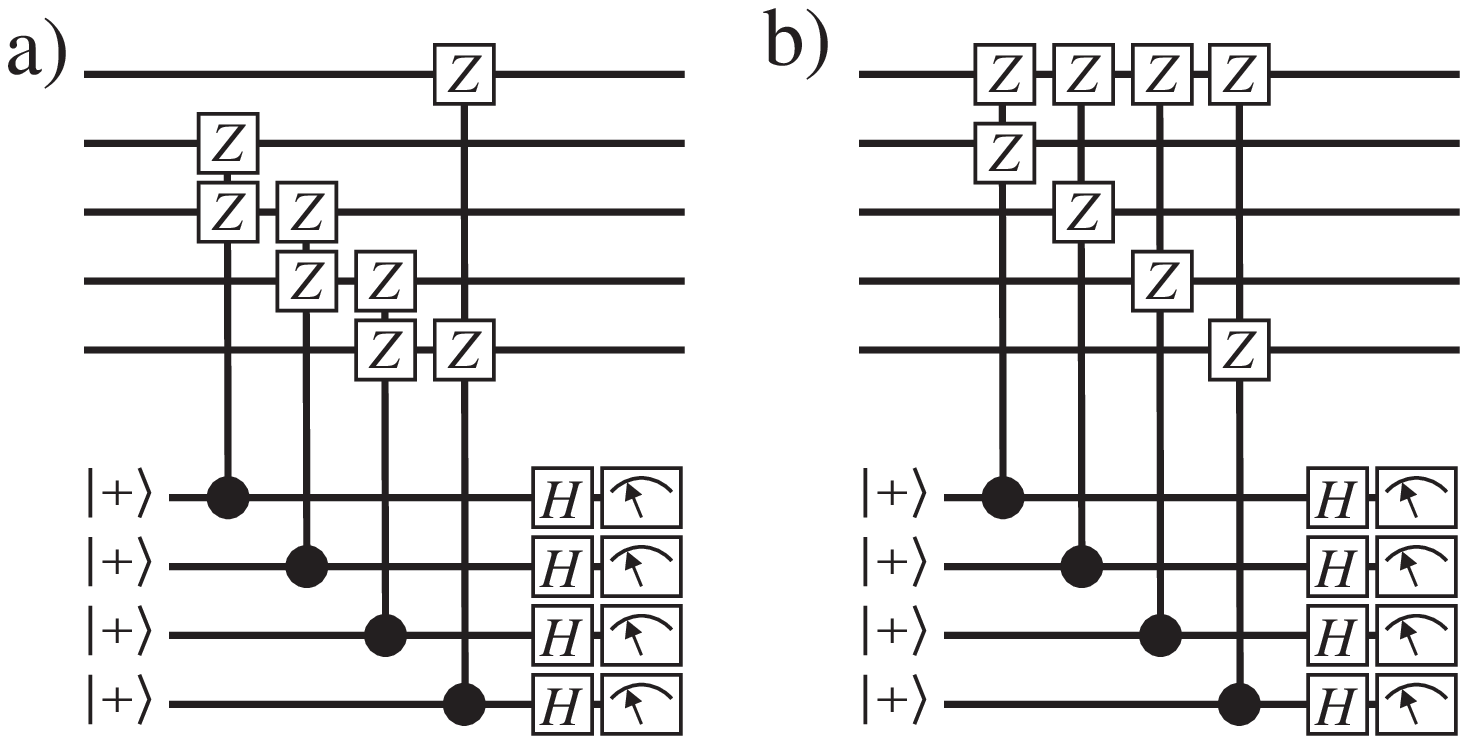}}}
\caption{An example of the verification construction. Both
networks show a complete set of parity check measurements to detect bit-errors in the codeword
$\ket{00000} + \ket{11111}$. (a) poorly constructed network; in several
places a single failure would allow a two-bit error to go undetected, therefore
repetition is needed. (b) well-constructed network: for any input ancilla state,
at the output the ancilla has uncorrelated $X$ errors (and may have correlated $Z$ errors)
whenever the check measurements all give zero.}
%\label{f_cav}
\end{figure}

Consider the syndrome given by a complete set of all the checks in
$H$, where each check is measured by preparing a single
verification qubit in $\ket{+} \equiv (\ket{0} +
\ket{1})/\sqrt{2}$, operating $\C Z$ gates from verification bit
to ancilla at locations given by a row of $H$, then measuring the
verification bit in the $\ket{+},\ket{-}$ basis. A sufficient condition
for CSS codes to achieve the desired behaviour is that
every syndrome of weight $w_{\rm s} \le t$ can be produced by
an error of weight $w_{\rm e} \le w_{\rm s}$.
If this holds then failure of $w = w_{\rm s}$ gates in
the verifier, allowing the error to pass unnoticed, will result in
an ancilla error of weight at most $w$.
This is true whether the error
was produced in the preparation stage or the verification stage or
a combination of both. We do not need to constrain the behaviour for
syndromes of weight $w_s > t$ because in any case a high-order failure
would be needed for the verifier to miss these. The sufficiency
of the condition relies on the following symmetry of the codeword.
Since we are verifying a single state (not a space), all errors are correctable.
This implies that different errors having the same syndrome must act identically
on the state, and indeed using standard stabilizer methods it is easy to
prove this\footnote{This is an example of degeneracy in quantum coding.}.
This symmetry is intimately related to the entanglement of the state. For a simple
example observe that $\ket{000} + \ket{111}$ is transformed in the same way
by errors $XII$ and $IXX$. This means that although a given syndrome $s$ is
produced by various errors, of which some have weight $> w_s$,
the effective error on the state in question is nevertheless of
weight $\le w_s$, as long as the condition holds.

We can satisfy the condition by exploiting basic properties of linear codes.
The parity checks form a linear vector space, and measurement of any spanning
set is enough to measure the whole space. We can therefore make a considered
choice of the spanning set. The property we need
is that no coset leader has higher weight than its syndrome.
To prove that this can always be arranged,
use the fact that the check matrix can always be written in
the standard form $H=(A,I)$ where $I$ is the $r \times r$ identity
matrix and $A$ is the rest of $H$ ($r$ is the number of rows in
$H$). We adopt this form; then the identity matrix part ensures
that all the syndromes of weight $w$ can be produced by errors of
weight $w$, because any error affecting only the last
$r$ bits gives a syndrome equal to the error, and all $2^r$
syndromes appear in this argument, QED.
Figure 1b shows a simple example.

We have now arranged that each parity check need only be measured once when
an ancilla is verified.
This shows that even a minimal QEC network, in the sense of one performing minimal
filtering of ancillas, can be fault-tolerant.
This reduction from $t+1$ repetitions to 1 permits both a large
reduction in the size of the computer and also an increase in the
tolerated memory noise, since with repeated verification $O(t+1)$
more ancillas would have to be prepared in parallel in order to
keep the overall recovery rate up, and each ancilla would also
have to survive $O(t+1)$ times longer before it can be used. A
large computation requires $t$ in the range $7$ (e.g.
$[[127,29,15]]$ BCH code) to $15$ (e.g. $[[7,1,3]]$ code
concatenated twice).

Repetition is still used to extract several copies of the
syndrome, to guard against correcting the computer on the basis of
a wrong syndrome. Next I reduce this also. Suppose
the data has an error $e$
whose true syndrome is $s$. If the syndrome were extracted bit by
bit, as in \cite{96:Shor,96:DiVincenzo}, then a single failure can
result in a single error $g$ in the {\em syndrome}. If $s+g$ is
accepted the `correction' of the data will leave the error $e+f$
in the data, where $f$ is the coset leader of the erroneous
syndrome $s+g$. $f$ is in general unrelated to $e$, therefore the
final state of the computer is liable to contain an error of
weight ${\rm wt}(e) + {\rm wt}(f)$ with probability $\sim
\epsilon^{{\rm wt(e)} + 1}$; the QEC soon breaks down since ${\rm
wt}(f)$ can be greater than 1.  However, in the method under
discussion, the syndrome is extracted indirectly by allowing the
error $e$ to propagate to the ancilla and then measuring the
ancilla. A failure which gives a single error $g$ in the ancilla
merely changes the error in the ancilla to $e+g$. Assuming this is
correctable, the interpretation of the ancilla measurement outcome
identifies $e+g$ as the error to be corrected in the data; the
final situation is then to leave the data with error $g$ with
probability $\sim \epsilon^{{\rm wt}(e) + 1}$, which is merely a
small addition to the probability of single errors in the data, so
is harmless. There remains a small need for syndrome extraction
repetition to guard against the comparatively few failure
locations that give larger weight $Z$ errors in the ancilla.

Next I will show how to compress the time required by the
preparation and verification ($H$) networks. I already noted that,
in order to be fault-tolerant, the $H$ network is constructed for
$H$ in the standard form $(A,I)$. This has the additional
advantage of almost minimizing the number of 1's in the matrix and
hence the number of gates in the verification network. The network
of $\C Z$ gates between verification bits and ancilla bits can be
parallelized to the degree that any group of $\C Z$ gates
involving different pairs of bits can take place simultaneously.
The problem of minimizing the number of time-steps is equivalent
to the problem of forming a latin rectangle the size of $A$, that
is $r \times (n-r)$, using an alphabet of minimal size, where the
places where $A$ has a zero need not be filled. Suppose we form
such a rectangle using integers from 1 to $N$, then each integer
gives the time-step in which the ancilla qubit of that column is coupled
to the verification bit for that row: the fact that no symbol
appears twice on a column guarantees that no ancilla bit is
involved in more than one gate at once; the fact that no symbol
appears twice on a row guarantees that no verifier bit is involved
in more than one gate at once. An example is given in figure 2. It
is clear that $N \ge w_{\rm max}$ where $w_{\rm max}$ is the
largest weight of a row or column of $A$, and it can be shown
from Hall's theorem in combinatorics that
a latin rectangle exists for this smallest possible $N$
\footnote{I am indebted to a referee for pointing this out.}.
The verification network is completed by a single
$\C Z$ from the verification bits to the last $r$ ancilla bits;
these can be simultaneous, so the total number of time-steps for
verification is $N+1+T_m$ where $T_m$ is the time required for
measurement of a set of qubits.

\begin{figure}[ht]
\centerline{\resizebox{!}{3.5 cm}{\includegraphics{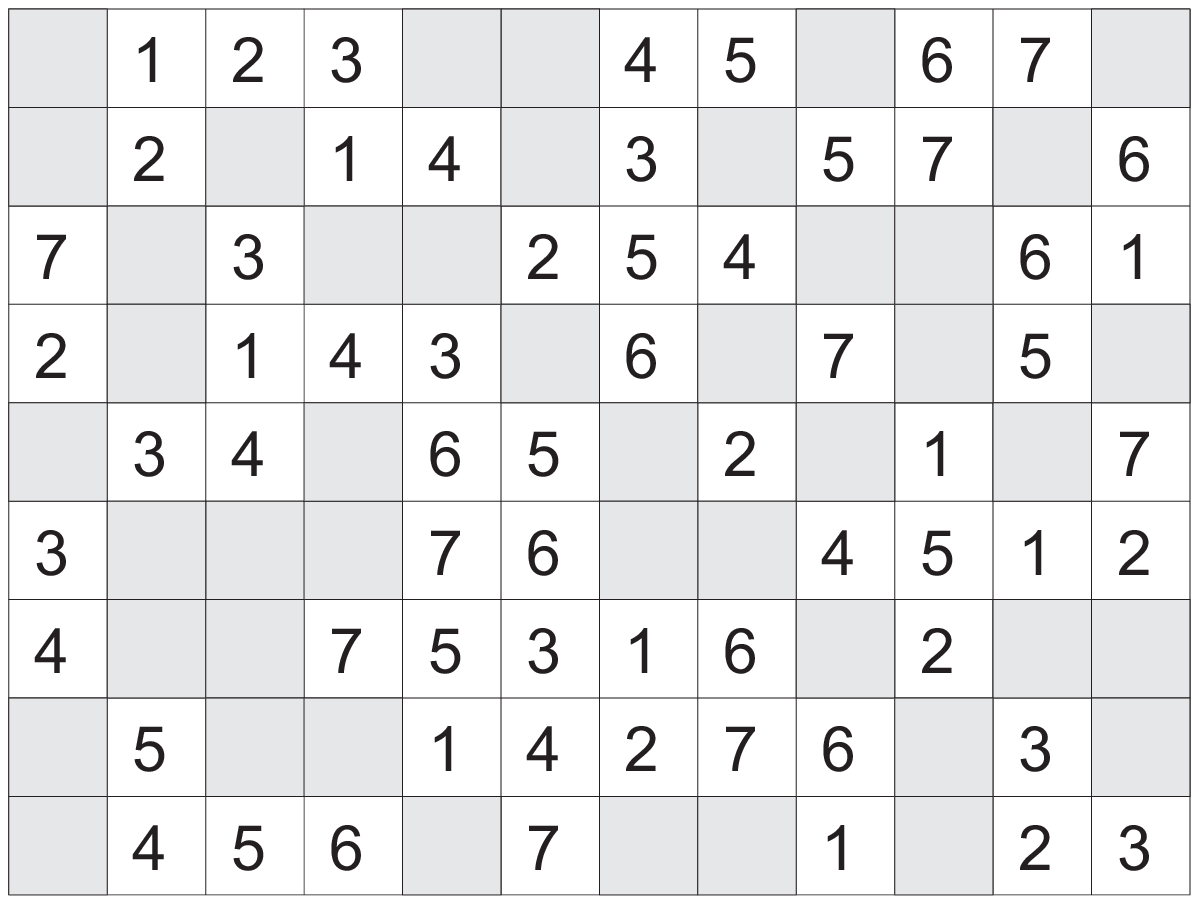}}}
\caption{Construction of the $G$ and $H$
networks, illustrated for the example of a $[[21,3,5]]$ code. The
$A$ matrix is shown, with blanks where zeros appear, and the other
locations numbered in a latin rectangle. The numbers indicate the
time step in which each controlled-gate is applied.}
%\label{f_cav}
\end{figure}

The method to prepare the $\ket{0}_L$ state need not involve
networks of gates, but if a network is used then the same analysis
shows that the $G$ network can be accomplished in $N$ time-steps
using $\C X$ gates (on bits prepared in $\ket{0}$ or $\ket{+}$).

To conclude, all CSS codes allow a network to prepare verified
ancillas which is both fault-tolerant and minimal; such networks
can also be compressed in time by a general procedure based on
latin squares. The results described have the common theme of
using structure in the design of the QEC network to serve to
enhance its ability to extract entropy from the computer. In
information theoretic terms, the structure is a form of
negative entropy in the ancillas, which allows them to
absorb more entropy from the data. The
practical result is that fewer checking operations and timesteps
are needed to run the QEC protocol, so that both a saving in
computer size and an increase in memory noise tolerated is
obtained. The saving is by a factor of the order of $t$, the
number of errors correctable by the code, which is in the range
approximately 7 to 15 for a large quantum computation.

\begin{acknowledgements}
This work was supported by the EPSRC and the Research Training and
Development and Human Potential Programs of the European Union.
\end{acknowledgements}

\bibliographystyle{apsrev}
\bibliography{quinforefs}

\end{document}